\documentclass[11pt]{article}
\usepackage[whole]{bxcjkjatype}
\usepackage[utf8]{inputenc}
\usepackage[
 a4paper,
 total={160mm,257mm},
 left=25mm,
 top=20mm]{geometry}
\usepackage{hyperref}
\usepackage{amsfonts}
\usepackage{amsmath}
\usepackage{amssymb}
\usepackage[mathscr]{eucal}
\usepackage{color}
\usepackage{authblk}
\usepackage{physics}
\usepackage{multicol}

\newcommand{\ee}{\mathrm{e}}
\newcommand{\ii}{\mathrm{i}}

\makeatletter

\@addtoreset{equation}{section}
\makeatother

\title{\textbf{ABCD of Kondo effect}} 
\author{\textsc{Taro Kimura}\\木村太郎}
\affil{Institut de Math\'ematiques de Bourgogne,\\ Universit\'e Bourgogne Franche-Comt\'e}
\date{}

\begin{document}

\maketitle

\begin{abstract}
    We explore the Kondo effect incorporating the localized impurity transforming under generic symmetry group $G$, that we call the $G$-Kondo effect.
    We derive the one-dimensional effective model coupled with the impurity, and studied the thermodynamic properties based on the boundary conformal field theory approach.
    We in particular study the impurity entropy at the conformal fixed point, and the low temperature behavior of the specific heat and the susceptibility.
    We also consider the Wilson ratio based on these results, and mention the case with exceptional group symmetry.
\end{abstract}

\tableofcontents
\vspace{1.5em}
\hrule

\section{Introduction}\label{sec:intro}

The Kondo effect (近藤効果) 
has been playing a central role in the study of strongly correlated electron systems, providing a lot of innovative ideas for our understanding of the quantum nature of many-body physics~\cite{Hewson:1993,Kondo:2009}.
It has been originally observed and studied in the magnetic impurity system, where the SU(2) spin degrees of freedom of the localized impurity spin interacting with the conduction electron plays a crucial role in the non-trivial behavior of the renormalization flow.
In fact, the Kondo effect seems the first experimental observation of the asymptotic freedom.

One of the interesting directions of the study of the Kondo effect is to incorporate higher-rank SU$(n)$ symmetry, discussed in the context of the quantum dot~\cite{Sasaki:2004PRL,Jarillo-Herrero:2005Nat}, ultracold atomic system~\cite{Nishida:2013PRL,Nishida:2016PRA}, heavy-flavor QCD~\cite{Yasui:2013xr,Hattori:2015hka}, etc.
In addition, there has been a recent proposal to implement the Majorana impurity, which realizes the SO($n$) Kondo system~\cite{Beri:2012tr,Altland:2013vza,Galpin:2014waa,Tsvelik:2014moa,Altland:2014ysa}.
See also~\cite{Lecheminant:2005PRL,Nonne:2011PRB,Stadler:2016PRB,Mitchell:2020so5} for further discussions on unconventional symmetries, including Sp$(n)$ symmetry.
Such developments in the Kondo effect motivates us to explore the quantum impurity system with generic symmetry group $G$, that we call the $G$-Kondo system.
In this paper, we mainly consider the classical groups, namely, \emph{ABCD of Kondo effect}:
\begin{align}
    A_n \ : \ \mathrm{SU}(n)
    \, , \qquad
    B_n \ : \ \mathrm{SO}(2n+1)    
    \, , \qquad
    C_n \ : \ \mathrm{Sp}(n)
    \, , \qquad
    D_n \ : \ \mathrm{SO}(2n)
    \, .
\end{align}
Here we apply the convention, s.t., $\mathrm{Sp}(1) = \mathrm{SU}(2)$, $\mathrm{Sp}(2) = \mathrm{SO}(5)$.
The key property of the Kondo effect is the non-perturbative behavior below the characteristic temperature, called the Kondo temperature $T_\text{K}$.
Such a behavior is universal, and observed in the system with arbitrary non-Abelian group $G$ in general:
For the Kondo coupling in the form of
\begin{align}
    \mathsf{H}_\text{K} = \lambda_\text{K} \sum_{a = 1}^{\dim G} \sigma^a S^a 
    \, ,
    \label{eq:K_coupling}
\end{align}
with the Kondo coupling $\lambda_\text{K}$, and the ``spin'' operators of the electron and the impurity, $\sigma^a$ and $S^a$ (the generators of the Lie algebra $\mathfrak{g} = \operatorname{Lie} G$), we obtain the scaling equation~\cite{Anderson:1970JPC} (See also \cite{Kogan:2019JPC}),
\begin{align}
    \dv{\lambda_\text{K}}{\log \Lambda} = - h^\vee \rho_0 \, \lambda_\text{K}^2
    \, ,
    \label{eq:scaling_eq}
\end{align}
where we define the energy scale $\Lambda$, the density of states $\rho_0$, and the dual Coxeter number $h^\vee$ (See Tab.~\ref{tab:h_hvee}).
The derivation of the scaling equation~\eqref{eq:scaling_eq} just uses the relation for the structure constant of the Lie algebra $\mathfrak{g} = \operatorname{Lie}G$ shown in~\eqref{eq:hvee}.

Since the electron-impurity coupling may be relatively small in the high temperature regime compared to $T_\text{K}$, we can apply the perturbative analysis there, whereas we should non-perturbatively explore the low temperature regime due to the asymptotic freedom property.
Conformal Field Theory (CFT) is one of the non-perturbative approaches to the low temperature regime of the Kondo system as demonstrated by Affleck--Ludwig~\cite{Affleck:1990zd,Affleck:1991tk,Affleck:1990iv,Affleck:1991yq,Affleck:1992ng,Ludwig:1994nf}.
See also review articles~\cite{Affleck:1995ge,Affleck:2008}.
In this framework, the impurity problem is replaced with the study of the boundary condition, which can be analyzed based on Boundary CFT (BCFT).

In this paper, we formulate the Kondo effect with generic symmetry group $G = ABCD$, and apply the BCFT approach in order to study the non-perturbative aspects of the $G$-Kondo system.
In Sec.~\ref{sec:formulation}, we first introduce the localized impurity which transforms under the symmetry group $G = ABCD$, and then formulate the quantum impurity problem based on the effective one-dimensional theory.
In particular, the impurity effect is described as the conformal boundary condition of the effective theory, which allows us to apply the BCFT techniques to the current problem.
Based on such a description, we analyze the thermodynamic properties of the $G$-Kondo system.
In Sec.~\ref{sec:g_factor}, we study the impurity contribution to the thermodynamic entropy (the impurity entropy), which would be a criterion whether the conformal fixed point is described as the Fermi liquid (FL) or the non-Fermi liquid (NFL).
In Sec.~\ref{eq:thermodynamics}, we then study the low temperature behaviors of the specific heat and the susceptibility using the conformal perturbation theory with the leading irrelevant operator.
We clarify the parameter condition in order to obtain the FL/NFL behavior.
We also mention the case with the exceptional group symmetry, $G = EFG$, in Sec.~\ref{sec:EFG}.

\section{Formulation of the impurity problem}\label{sec:formulation}

The Kondo problem is originally formulated in three dimensions with a conduction electron and a localized impurity.
Assuming the electron-impurity interaction takes a form of the $\delta$-function, we can focus on the $s$-wave scattering, and the system is effectively reduced to the one-dimensional model defined on a semi-infinite domain $x \in \mathbb{R}_+$.
Furthermore, since we are interested in the excitation in the vicinity of the Fermi energy, we approximate the dispersion relation by $\epsilon(k) \approx v_\text{F}(k - k_\text{F})$, which leads to the effective Hamiltonian,
\begin{align}
    \mathsf{H} 
    & = \frac{v_\text{F}}{2\pi} \int_0^\infty \dd{x}
    \qty[ 
    \psi_{\text{L},i\alpha}^\dag(x) \ii \partial_x \psi_{\text{L},i\alpha}(x) 
    - \psi_{\text{R},i\alpha}^\dag(x) \ii \partial_x \psi_{\text{R},i\alpha}(x) 
    ]
    + \mathsf{H}_\text{K}
    \, ,
    \label{eq:Ham1}
\end{align}
where $v_\text{F}$ is the Fermi velocity, and $\psi_{\text{L}/\text{R},i\alpha}$ is the left/right moving fermion with the spin and channel (flavor) indices, $\alpha = 1,\ldots,n$, $i = 1,\ldots,k$.
The electron-impurity interaction is described by $\mathsf{H}_\text{K}$ defined in \eqref{eq:K_coupling}.

\subsection{Non-Abelian bosonization}

For the complex (charged) fermion, the Hamiltonian~\eqref{eq:Ham1} is invariant under the following transformations,
\begin{align}
    G = \mathrm{SU}(n) : \ \psi_{\alpha} \mapsto (g\psi)_\alpha
    \, , \quad
    G' = \mathrm{SU}(k) : \ \psi_{i} \mapsto (g'\psi)_i
    \, , \quad
    \mathrm{U}(1) : \ \psi \mapsto \ee^{\ii \phi} \psi
    \, .
\end{align}
Namely, the fermion operator transforms in the bifundamental representation of $G \times G'$.
Assuming the interaction $\mathsf{H}_\text{K}$ is insensitive (singlet) to the channel symmetry $G'$, we may consider the right mover $\psi_\text{R}$ as the analytic continuation of the left mover to the negative region, $\psi_{\text{L},\alpha}(x) = (g \psi_\text{R})_\alpha(-x) =: \psi_{\alpha}(x)$ with $g \in G$.
Then, we obtain the chiral expression of the Hamiltonian~\eqref{eq:Ham1} as follows:
\begin{align}
    \mathsf{H}
    & =
    \frac{v_\text{F}}{2\pi} \int_{-\infty}^\infty \dd{x}
    \psi_{i\alpha}^\dag(x) \ii \partial_x \psi_{i\alpha}(x) 
    + \mathsf{H}_\text{K}    
    \, .
    \label{eq:Ham2}    
\end{align}
The interaction term, originally introduced in~\eqref{eq:K_coupling}, is rewritten as
\begin{align}
    \mathsf{H}_\text{K}
    = \lambda_\text{K} \sum_{a = 1}^{\dim G} \psi_{i\alpha}^\dag (x) (t^a)_{\alpha\beta} \psi_{i\beta}(x) \, S^a \, \delta(x)
    = \lambda_\text{K} \, J^a(x) \, S^a \, \delta(x)
    \, ,
\end{align}
where the localized impurity spin operator is denoted by $S^a \, \delta(x)$ for $a = 1,\ldots,\dim G$.
We define the current operators with respect to $(G,G',\mathrm{U}(1))$-symmetry (non-Abelian bosonization; See, for example,~\cite{Gogolin:2004rp} for details):
\begin{align}
    J^a(x) = {: \psi^\dag_{i\alpha} (t^a)_{\alpha\beta} \psi_{i\beta} :}(x)
    \, , \quad
    J^A(x) = {: \psi^\dag_{i\alpha} (T^A)_{ij} \psi_{j\alpha} :}(x)
    \, , \quad
    J(x) = {: \psi^\dag_{i\alpha} \psi_{i\alpha} :}(x)
\end{align}
where $(t^a)_{a = 1,\ldots,\dim G}$, $(T^A)_{A = 1,\ldots,\dim G'}$ are the generators of the corresponding Lie algebras, $\mathfrak{g} = \operatorname{Lie} G$ and $\mathfrak{g}' = \operatorname{Lie} G'$, and the normal ordering symbol denoted by $: \mathcal{O} :$.
The levels of these currents are respectively given by $k$, $n$, and $kn$.
The level-$k$ $G$-symmetry current ($G_k$-current for short) obeys the operator product expansion (OPE),
\begin{align}
    J^a(z) J^b(w) = \frac{\frac{k}{2} \, \delta^{ab}}{(z - w)^2} + \frac{f^{ab}_{~~c} \, J^c(w)}{z - w} + \cdots
    \label{eq:OPE_JJ}
\end{align}
where $(f^{ab}_{~~c})$ is the structure constant of the Lie algebra $\mathfrak{g} = \operatorname{Lie} G$.
Applying the mode expansion, 
\begin{align}
    J_n^a = \oint \frac{\dd{z}}{2 \pi \ii} \, z^n \, J^a(z)
    \ \iff \
    J^a(z) = \sum_{n \in \mathbb{Z}} J_n^a \, z^{-n-1}
    \, ,
\end{align}
the OPE is equivalent to the algebraic relation of the affine Lie algebra $\widehat{\mathfrak{g}}_k$, which is an infinite dimensional extension of $\mathfrak{g}$,
\begin{align}
    \comm{J_n^a}{J_m^b}
    = f^{ab}_{~~c} J_{n+m}^c + \frac{k}{2} n \, \delta^{ab} \delta_{n+m,0}
    \, .
\end{align}
The same argument is applied to the $G'$-current.

We put $v_\text{F} = 1$ for simplicity.
Then, the Hamiltonian~\eqref{eq:Ham2} is rewritten in the quadratic form of the current operators (Sugawara form),
\begin{align}
    \mathsf{H} 
    = 
    \frac{1}{2 \pi}
    \qty[
    {\frac{1}{n + k} : J^a J^a :}(x) 
    + {\frac{1}{k + n} : J^A J^A :}(x)
    + {\frac{1}{2 kn} :JJ:}(x)
    ]
    + \lambda_\text{K} J^a S^a \delta(x)
    \, .
    \label{eq:Ham3}
\end{align}
Such a decomposition of the $kn$-component complex fermion system is a consequence of the conformal embedding
\begin{align}
    \widehat{\mathfrak{su}(n)}_k 
    \oplus \widehat{\mathfrak{su}(k)}_n
    \oplus \widehat{\mathfrak{u}(1)}_{kn}
    \subset
    \widehat{\mathfrak{u}(kn)}_{1}
    \, ,
    \label{eq:embedding1}
\end{align}
which corresponds to the spin-charge(-channel) separation of the one-dimensional fermion.
We see the agreement of the central charges
\begin{align}
    c\qty[\widehat{\mathfrak{su}(n)}_k]
    + c\qty[\widehat{\mathfrak{su}(k)}_n]
    + c\qty[\widehat{\mathfrak{u}(1)}_{kn}]
    & =
    \frac{k(n^2 - 1)}{k + n}
    + \frac{n(k^2 - 1)}{n + k}
    + 1
    \nonumber \\
    & = kn
    \nonumber \\    
    & = c\qty[\widehat{\mathfrak{u}(kn)}_{1}]
    \, ,
    \label{eq:cc1}
\end{align}
where the central charge of the affine Lie algebra $\widehat{\mathfrak{g}}_k$ is given by
\begin{align}
    c\qty[ \widehat{\mathfrak{g}}_k ] = \frac{k \dim \mathfrak{g}}{k + h^\vee}
    \label{eq:cc}
\end{align}
with $h^\vee$ the dual Coxeter number of $\mathfrak{g}$, e.g., $h^\vee = n$ for $\mathfrak{su}(n)$.
See Appendix~\ref{sec:Lie_alg} for details.

In fact, the current operator obeys the same OPE~\eqref{eq:OPE_JJ} under the redefinition,
\begin{align}
    J^a(x) \ \longrightarrow \ J^a(x) + 2 \pi \delta(x) S^a 
    \, .
    \label{eq:current_shift}
\end{align}
Therefore, if the coupling constant takes the critical value $\lambda_\text{K}^* = 2/(n + k)$, the interaction term in the Hamiltonian~\eqref{eq:Ham3} is absorbed into the quadratic term,
\begin{align}
    \mathsf{H} 
    = 
    \frac{1}{2\pi}
    \qty[
    {\frac{1}{n + k} : J^a J^a :}(x) 
    + {\frac{1}{k + n} : J^A J^A :}(x)
    + {\frac{1}{2 kn} :JJ:}(x)
    ]
    \, ,
\end{align}
where we omit the non-dynamical term $S^a S^a$.
In fact, the critical value $\lambda_\text{K}$ is interpreted as the value of the coupling constant at the conformal fixed point, and the impurity effect would be incorporated as the conformal boundary condition labeled by the representation $R_\text{imp}$ under the symmetry group~$G$.

\subsection{Real fermion}\label{sec:real_embed}

We can similarly formulate the impurity problem with the real (Majorana) fermions, by replacing the complex conjugate $\psi^\dag$ with $\psi$.
We have seen that, the conformal embedding~\eqref{eq:embedding1} plays a crucial role to specify the symmetry of the system.
In the case with the real fermions, we instead have the following embeddings~\cite{Hasegawa:1989PRIMS,Verstegen:1990at}:
\begin{subequations}\label{eq:embedding_real}
\begin{align}
    \widehat{\mathfrak{so}(n)}_k 
    \oplus \widehat{\mathfrak{so}(k)}_n
    & \subset
    \widehat{\mathfrak{so}(kn)}_{1}   
    \quad \text{for} \quad
    kn : \ \text{even}
    \, ,
    \\
    \widehat{\mathfrak{sp}(n)}_k 
    \oplus \widehat{\mathfrak{sp}(k)}_n
    & \subset
    \widehat{\mathfrak{so}(4kn)}_{1}
    \, .
\end{align}
\end{subequations}
The central charges are given by
\begin{subequations}\label{eq:cc_real}
\begin{align}
    c\qty[ \widehat{\mathfrak{so}(n)}_k ]
    + c\qty[ \widehat{\mathfrak{so}(k)}_n ]
    & 
    = \frac{k}{2} \frac{n^2 - n}{n + k - 2}
    + \frac{n}{2} \frac{k^2 - k}{k + n - 2}
    = \frac{kn}{2}
    = c\qty[ \widehat{\mathfrak{so}(kn)}_1 ]
    \, , \\
    c\qty[ \widehat{\mathfrak{sp}(n)}_k ]
    + c\qty[ \widehat{\mathfrak{sp}(k)}_n ]
    & 
    = \frac{kn(2n+1)}{k + n + 1}
    + \frac{nk(2k+1)}{n + k + 1}
    = 2kn
    = c\qty[ \widehat{\mathfrak{so}(4kn)}_1 ]
    \, .
\end{align}
\end{subequations}
The latter is also interpreted as the embedding for the $kn$-component quaternionic fermion.
Based on these embeddings, we have two options for the symmetry groups for the real fermion system:
\begin{align}
    (G,G') =
    \begin{cases}
    (\mathrm{SO}(n), \mathrm{SO}(k)) \\
    (\mathrm{Sp}(n), \mathrm{Sp}(k)) \\
    \end{cases}
\end{align}

We then consider the Sugawara Hamiltonian for the real fermion system:
\begin{align}
    \mathsf{H} = \frac{1}{2 \pi}
    \qty[
    {\frac{1}{k + h^\vee(\mathfrak{g})} :J^a J^a:}(x)
    + {\frac{1}{n + h^\vee(\mathfrak{g}')} :J^A J^A:}(x)
    ]
    + \lambda_\text{K} J^a S^a \delta(x)
    \, .
\end{align}
where $S^a$ is the impurity ``spin'' which transforms under $G = \mathrm{SO}(n)$ and Sp$(n)$, respectively.
See Tab.~\ref{tab:h_hvee} for the list of the dual Coxeter number.
We remark that, compared to the complex fermion system~\eqref{eq:Ham3}, there is no $\mathrm{U}(1)$-current (the charge current) in this case.
This is because the analog of $\mathrm{U}(1)$ in this case is the discrete group $\mathrm{O}(1) = \mathbb{Z}_2$.
Precisely speaking, we should consider the spin group (the universal cover of SO group) in the conformal embedding, depending on the parity of the rank.
See, for example,~\cite{Aharony:2016jvv} for details.

If the coupling constant takes the critical value $\lambda_\text{K}^* = 2/(k + h^\vee(\mathfrak{g}))$, the interaction term is combined into the quadratic form by the $G_k$-current shift as in~\eqref{eq:current_shift},
\begin{align}
    \mathsf{H} = \frac{1}{2 \pi}
    \qty[
    {\frac{1}{k + h^\vee(\mathfrak{g})} :J^a J^a:}(x)
    + {\frac{1}{n + h^\vee(\mathfrak{g}')} :J^A J^A:}(x)
    ]
    \, .
\end{align}
Again, we consider the conformal boundary condition characterized by the representation $R_\text{imp}$ of the impurity under the symmetry group $G$.

\section{Impurity entropy}\label{sec:g_factor}

The impurity contribution to the entropy (impurity entropy; also called the boundary entropy) provides a useful criterion for the FL/NFL behavior at the conformal fixed point.
In this Section, we compute the impurity entropy of $G$-Kondo system based on the CFT formalism discussed in Sec.~\ref{sec:formulation}.

The impurity entropy of the Kondo system is obtained through the so-called $g$-factor, which counts the boundary degeneracy~\cite{Affleck:1991tk}:
\begin{align}
    S_\text{imp} = \log g(R_\text{imp})
\end{align}
where we denote the representation of the impurity under the symmetry group $G$ by $R_\text{imp}$.
If the $g$-factor takes an integral value, $g(R) \in \mathbb{N}$, the fixed point is supposed to be the FL, since one can ``count'' the degeneracy of the impurity states, and the original picture which consists of the conduction electron and the localized impurity still holds at the fixed point.
On the other hand, it would be described as the NFL if the $g$-factor becomes an irrational number, $g(R) \not\in \mathbb{N}$, where the original separated picture of the electron and the impurity is not available any longer.
We remark that there are several exceptional situations for the NFL where the corresponding $g$-factor takes an integral value due to an accidental symmetry.

The $g$-factor is given by the modular $\mathsf{S}$-matrix of the affine Lie algebra $\widehat{\mathfrak{g}}_k$ denoted by $\mathsf{S}_{RR'}$~\cite{DiFrancesco:1997nk}:%
\begin{align}
    g(R) 
    = \frac{\mathsf{S}_{0R}}{\mathsf{S}_{00}}
    = \chi_R \qty( q^\rho )
    \, ,
    \label{eq:g_fac}
\end{align}
where $\chi_R(X)$ is the character of the representation $R$ of the Lie algebra $\mathfrak{g}$ evaluated with the Cartan torus $X \in (\mathbb{C}^\times)^{\operatorname{rk} \mathfrak{g}}$, and $\rho$ is the Weyl vector of $\mathfrak{g}$ given by \eqref{eq:Weyl_vec}.
The character formula is summarized in Appendix~\ref{sec:character} for $G = ABCD$.
The parameter $q \in \mathbb{C}^\times$, $\abs{q} = 1$, is defined with the level and the dual Coxeter number $(k,h^\vee)$ as follows:
\begin{align}
    q = \exp \left( \frac{2 \pi \ii}{k + h^\vee} \right)
    \, .
\end{align}
We remark that the expression of the $g$-factor~\eqref{eq:g_fac} coincides with the unknot Wilson loop expectation value of $G_k$ Chern--Simons theory on the three-sphere $\mathbb{S}^3$~\cite{Witten:1988hf}.
See also~\cite{Bachas:2004sy,Gaiotto:2020fdr,Gaiotto:2020dhf} for further discussion on the relation to the line defect.

In fact, the $g$-factor is given as the principal specialization of the character of the representation $R$, a.k.a., the quantum dimension of the representation $R$,
\begin{align}
    g(R) = \dim_q R \ge 1
    \, .
\end{align}
The quantum dimension is reduced to the ordinary dimension in the limit $q \to 1$, $\lim_{q \to 1} \dim_q R = \dim R \in \mathbb{N}$, which is interpreted as the $g$-factor in the weak coupling regime.
In addition, since it is a polynomial of $(q,q^{-1})$ with non-negative integer coefficients, it is smaller than the classical value, $g(R) \le \dim R$.
This implies that the $g$-factor decreases (possibly, monotonically) along the renormalization flow to the low-energy regime.
This is called the $g$-theorem, a boundary analog of the $c$-theorem in CFT.

We remark that, in the large $k$ limit, the asymptotic behavior of the impurity entropy is given by
\begin{align}
    S_\text{imp} 
    \ \xrightarrow{k \to \infty} \
    \log \dim R
    \, .
\end{align}
Hence, the $g$-factor takes an integral value, and thus the the fixed point is described as the FL in the limit $k \to \infty$.
This implies that the quantum correction to the $g$-factor is largely suppressed in in this case.

\subsection{SU($n$) theory}

We consider the $g$-factor for $G = \mathrm{SU}(n)$ theory.
Let us consider the impurity, which transforms in the $n$-dimensional (natural) representation of $\mathrm{SU}(n)$ denoted by $\mathbf{n} = R(1,0,\ldots,0)$.
Then, applying the formula shown in Appendix~\ref{sec:character}, we obtain the $g$-factor as follows:
\begin{align}
    g(\mathbf{n}) 
    = \frac{q^{\frac{1}{2}(n - 1)} - q^{-\frac{1}{2}(n - 1)}}{q^{\frac{1}{2}} - q^{-\frac{1}{2}}}
    = q^{\frac{1}{2}(n - 1)} + q^{\frac{1}{2}(n - 3)} + \cdots + q^{-\frac{1}{2}(n - 1)}
    \ \xrightarrow{q \to 1} \ 
    n
    \, .
\end{align}
We remark that its conjugate representation gives rise to the same $g$-factor, $g(\bar{\mathbf{n}}) = g(\mathbf{n})$.
The numerical values of the $g$-factor with several $(n,k)$ are given as follows~\cite{Kimura:2016zyv}:
 \begin{align}
 \begin{tabular}{c@{\qquad}ccccc@{\qquad}c}\hline\hline
  $n$ & $k=1$ & $k=2$ & $k=3$ & $k=4$ & $k = 5$ & $k=\infty$\\ \hline
  2 & 1 & 1.4142... & 1.6180... & 1.7320... & 1.8019... & 2 \\
  3 & 1 & 1.6180... & 2 & 2.2469... & 2.4142... & 3 \\
  4 & 1 & 1.7320... & 2.2469... & 2.6131... & 2.8793... & 4\\
  5 & 1 & 1.8019... & 2.4142... & 2.8793... & 3.2360... & 5\\[.3em]
  $\infty$ & 1 & 2 & 3 & 4 & 5 & $\infty$ \\[.3em]
  \hline\hline
 \end{tabular}
 \label{tab:g_A}
  \end{align}
We see that the $g$-factor takes irrational values for generic $(n,k)$, except at $k = 1, \infty$, $n \to \infty$.
This implies that it behaves as the FL at large $(n,k)$ and at $k = 1$, while it would be described as the NFL in other cases.
(The case with $(n,k) = (3,3)$ is exceptional.)
The symmetry $n \leftrightarrow k$ is a consequence of the level-rank duality of the affine Lie algebras, $\widehat{\mathfrak{su}(n)}_k$ and $\widehat{\mathfrak{su}(k)}_n$, which is also related to the conformal embedding~\eqref{eq:embedding1}~\cite{Altschuler:1989nm,Fuchs:1989rv,Naculich:1990pa}.

\subsection{SO($n$) theory}

We consider the $n$-dimensional vector representation and the spinor representation of $\mathrm{SO}(n)$ theory.

\subsubsection{Vector representation}

The $g$-factor for the vector representation of $\mathrm{SO}(n)$ denoted by $\mathbf{n} = R(1,\ldots,0)$ is given by
\begin{align}
 g(\mathbf{n})
 & =
 \begin{cases}
  \displaystyle
  q^{m - 1/2} + \cdots + q^{1/2} + 1 + q^{-1/2} + \cdots + q^{-m+1/2}
  &
  (n = 2m + 1) \\
  q^{m - 1} + \cdots + q + 2 + q^{-1} + \cdots + q^{-m+1}
  &
  (n = 2m)  
 \end{cases}
 \nonumber \\
 & \xrightarrow{q \to 1} \ n
 \, ,
\end{align}
and the numerical values are given as follows:
 \begin{align}
 \begin{tabular}{c@{\qquad}ccccc@{\qquad}c}\hline\hline
  $n$ & $k=1$ & $k=2$ & $k=3$ & $k=4$ & $k=5$ & $k=\infty$\\ \hline
  3 & 1 & 2 & 2.4142... & 2.6180... & 2.7320... & 3 \\
  4 & 1 & 2 & 2.6180... & 3 & 3.2469... & 4 \\
  5 & 1 & 2 & 2.7320... & 3.2469... & 3.6131... & 5\\
  6 & 1 & 2 & 2.8019... & 3.4142... & 3.8793... & 6\\[.3em]
  $\infty$ & 1 & 2 & 3 & 4 & 5 & $\infty$ \\[.3em]
  \hline\hline
 \end{tabular}
 \label{tab:g_BD}
  \end{align}
In this case, the $g$-factor becomes $g = 1, 2$ for $k = 1, 2$, and it is expected to be described as the FL.
The remaining cases would be the NFL.
We can see the level-rank duality between $\widehat{\mathfrak{so}(n)}_k$ and $\widehat{\mathfrak{so}(k)}_n$ similarly to SU($n$) theory.
We remark that $\mathrm{SO}(4)_4$ exceptionally gives an integral value due to the isomorphism, $\mathrm{SO}(4) = \mathrm{SU}(2) \times \mathrm{SU}(2)$. 

\subsubsection{Spinor representation}

We then study the spinor representation, $\mathscr{S} = R\qty( \frac{1}{2},\ldots,\frac{1}{2})$, whose dimension is $2^{h/2}$ with the Coxeter number $h$ (Tab.~\ref{tab:h_hvee}):
It is given by $2^{m}$ and $2^{m-1}$ for SO($2m+1$) and SO($2m$).
Precisely speaking, SO$(2m)$ theory has two spin representations, $\mathscr{S}^+$ and $\mathscr{S}^-$, labeled by $\mathscr{S}^\pm = R\qty( \frac{1}{2},\ldots,\pm \frac{1}{2})$, which give rise to the same $g$-factor.

The numerical values of the $g$-factor are given by
\begin{align}
 \begin{tabular}{c@{\qquad}ccccc@{\qquad}c}\hline\hline
  $n$ & $k=1$ & $k=2$ & $k=3$ & $k=4$ & $k=5$ & $k=\infty$\\ \hline
  3 & 1.4142... & 1.7320... & 1.8477... & 1.9021... & 1.9318... & 2 \\
  4 & 1 & 1.4142... & 1.6180... & 1.7320... & 1.80194... & 2 \\
  5 & 1.4142... & 2.2360... & 2.7320... & 3.0489... & 3.2619... & 4 \\
  6 & 1 & 1.7320... & 2.2469... & 2.6131... & 2.8793... & 4 \\[.3em]
  \hline\hline
 \end{tabular}
 \label{tab:g_BD_sp}
\end{align}
We remark that the spinor representation of SO(6) is equivalent to the $\mathbf{4}$-representation of SU(4) as in \eqref{tab:g_A} as a consequence of the isomorphism between SO(6) and SU(4).

In particular, we see the peculiar behavior,
$g(\mathscr{S}) = 1$ for SO$(2m)_{k=1}$ and $g(\mathscr{S}) = \sqrt{2} = 1.4142...$ for SO$(2m+1)_{k=1}$.
This behavior is consistent with the analysis in~\cite{Tsvelik:2014moa} studying the $\mathrm{SO}(n)_{k = 1}$ impurity system:
The Majorana impurity system would be in general described as the NFL except for $k=1$ of SO($2m$) theory.
In fact, the central charges of SO($n$) theories are given by
\begin{subequations}
\begin{align}
    c \left[ \widehat{\mathfrak{so}(2m)}_k \right]
    & = \frac{km(2m-1)}{k + 2m - 2}
    \ \xrightarrow{k \to 1} \
    m 
    \, , \\
    c \left[ \widehat{\mathfrak{so}(2m+1)}_k \right]
    & = \frac{km(2m+1)}{k + 2m - 1}
    \ \xrightarrow{k \to 1} \
    m + \frac{1}{2}
    \, .
\end{align}
\end{subequations}
This is interpreted as follows:
For the SO$(2m)$ theory, there are $2m$ Majorana fermions, and they make pairs to form $m$ complex fermions to be described as the FL.
In the case of SO$(2m+1)$ theory, on the other hand, $2m+1$ Majorana fermions similarly gives rise to $m$ pairs ($m$ complex fermions) and a single Majorana fermion.
The factor $\frac{1}{2}$ is actually interpreted as the contribution of the Majorana fermion, which leads to the NFL behavior.

\subsection{Sp($n$) theory}\label{sec:g_Sp}

Let us turn to the $\mathrm{Sp}(n)$ theory.
In this case, we consider the $2n$-dimensional (natural) representation of $\mathrm{Sp}(2n)$, $\mathbf{2n} = R(1,0,\ldots,0)$.
The $g$-factor is given as follows:
\begin{align}
    g(\mathbf{2n}) = q^n + \cdots + q + q^{-1} + \cdots q^{-n}
    \ \xrightarrow{q \to 1} \ 
    2 n
    \, .
\end{align}
The numerical values of the $g$-factor are given as follows:
\begin{align}
 \begin{tabular}{c@{\qquad}ccccc@{\qquad}c}\hline\hline
  $n$ & $k=1$ & $k=2$ & $k=3$ & $k=4$ & $k = 5$ & $k=\infty$\\ \hline
  1 & 1 & 1.4142... & 1.6180... & 1.7320... & 1.8019... & 2 \\  
  2 & 1.4142... & 2.2360... & 2.7320... & 3.0489... & 3.2619... & 4 \\
  3 & 1.6180... & 2.7320... & 3.4939... & 4.0273... & 4.4114... & 6 \\
  4 & 1.7320... & 3.0489... & 4.0273... & 4.7587... & 5.3137... & 8 \\[.3em]
  $\infty$ & 2 & 4 & 6 & 8 & 10 & $\infty$ \\[.3em]
  \hline\hline
 \end{tabular}
\end{align}
In order to obtain consistent results for Sp$(n)$ theory, we have to rescale the $q$-parameter as $q \to q^{1/2}$, verified with the isomorphisms, $\mathrm{Sp}(1) = \mathrm{SU}(2)$ and $\mathrm{Sp}(2) = \mathrm{SO}(5)$. 
We can see the level-rank duality between $\widehat{\mathfrak{sp}(n)}_k$ and $\widehat{\mathfrak{sp}(k)}_n$ as well.
The limit values are $\displaystyle \lim_{n \to \infty} g = 2k$ and $\displaystyle \lim_{k \to \infty} g = 2n$.
We remark that the cases with $n = 1$ and $n = 2$ are equivalent to SU(2) and (the spin representation of) SO(5) shown in \eqref{tab:g_A} and \eqref{tab:g_BD_sp}, respectively, due to the isomorphisms.

A specific feature of Sp($n$) theory is that the $g$-factor becomes irrational even for $k = 1$ (except for $n = 1$, which is isomorphic to SU(2)).
Therefore, the Sp($n$) fixed point would be described as the NFL in general.
The central charge of the Sp($n$) theory is given by
\begin{align}
    c\qty[ \widehat{\mathfrak{sp}(n)}_k ]
    = \frac{kn(2n+1)}{k + n + 1}
    \ \xrightarrow{k \to 1} \ 
    \frac{n (2n+1)}{n + 2}
    \, .
\end{align}
In fact, it is a non-integer value even at $k = 1$, which also suggests that the $\mathrm{Sp}(n)_{k = 1}$ theory would behave as the NFL.

\section{Low temperature thermodynamics}\label{eq:thermodynamics}

In addition to the conformal fixed point corresponding to zero temperature $T = 0$, we can also study the low temperature behavior of the thermodynamic quantities based on the conformal perturbation theory.

\subsection{Bulk contribution}

We first consider the bulk contribution to the thermodynamic quantities, the specific heat and the susceptibility~\cite{Bloete:1986qm,Affleck:1986bv,Affleck:1986sc}:
\begin{align}
    C_\text{bulk} = \frac{\pi}{3} c \, T
    \, , \qquad
    \chi_\text{bulk} = \frac{k}{2 \pi}
    \, ,
\end{align}
where $c$ is the total central charge, corresponding to the conformal embeddings, \eqref{eq:cc1} and \eqref{eq:cc_real},
\begin{align}
    c = bkn =
    \begin{cases}
    kn & (\mathrm{SU}(n)_k) \\[.5em]
    \displaystyle \frac{kn}{2} & (\mathrm{SO}(n)_k) \\
    2 kn & (\mathrm{Sp}(n)_k)
    \end{cases}
    \label{eq:c_tot}
\end{align}
with the symmetry parameter 
\begin{align}
    \mathrm{SU}(n) : \
    b = 1 
    \, , \qquad
    \mathrm{SO}(n) : \
    b = \frac{1}{2} 
    \, , \qquad
    \mathrm{Sp}(n) : \
    b = 2 
    \, .
    \label{eq:b_parameter}
\end{align}
We remark that the parameter $\beta = 2 b$ is called the Dyson index in the context of random matrix theory.
See~\cite{Eynard:2015aea} for details.

The susceptibility considered here is with respect to the external $G$-field (magnetic field), which is coupled to the $G$-current in the form of $h^a J^a$.
The $G$-symmetry allows us to take the specific direction for the coupling term $h^1 J^1 =: h J^1$ without loss of generality. 
See~\cite{Beri:2012tr,Altland:2013vza,Tsvelik:2014moa,Altland:2014ysa} for the proposal to realize the magnetic field in $\mathrm{SO}(n)$ Kondo system.

\subsection{Impurity contribution}\label{eq:therm_imp}

In order to compute the finite temperature behavior, we apply the perturbative analysis with the leading irrelevant operators as follows~\cite{Affleck:1990zd,Affleck:1990iv}:
\begin{subequations}
\begin{align}
    \mathcal{O}_1 & = {:J^a J^a:(x)} \, \delta(x)
    \, , \\
    \mathcal{O}_2 & = {:J_{-1}^a \phi^a:(x)} \, \delta(x)
    \, ,
\end{align}
\end{subequations}
with
\begin{align}
    \delta \mathsf{H} = - \lambda_{\sigma} \, \mathcal{O}_{\sigma}
    \qquad (\sigma = 1, 2)
    \, .
\end{align}
The first operator $\mathcal{O}_1$ is the quadratic operator with the dimension $\Delta_1 = 2$.
Therefore, the dimensional analysis shows that the corresponding coupling constant $\lambda_1$ should be proportional to $T_\text{K}^{-1}$.

The second operator $\mathcal{O}_2$ is the first descendent of the adjoint primary operator $\phi^a(x)$ with the (conformal) dimension~\cite{Knizhnik:1984nr},
\begin{align}
    \Delta_\phi = \frac{h^\vee}{k + h^\vee}
    \, ,
    \label{eq:adj_dim}
\end{align}
for $\widehat{\mathfrak{g}}_k$ theory, where we apply the conformal dimension formula of the primary operator associated with the representation $R$ of the Lie algebra $\mathfrak{g}$, $\Delta_R = C_2(R)/(k + h^\vee)$.
$C_2(R)$ is the quadratic Casimir of the representation $R$.
For the adjoint representation, it is given by the dual Coxeter number \eqref{eq:hvee}.

The dimension of the second operator becomes $\Delta_2 = 1 + \Delta_\phi$, and the corresponding coupling constant $\lambda_2$ should be proportional to $T_\text{K}^{-\Delta_\phi}$.
We remark that this operator $\mathcal{O}_2$ may be identified with the so-called $\hat{h}$-scalar in the context of the coset CFT~\cite{Bowcock:1988vs}.
Since $\Delta_\phi < 1$ for $k \ge 1$, and thus $\Delta_2 < \Delta_1$, the second operator $\mathcal{O}_2$ is the leading irrelevant operator if it exists.
In fact, the adjoint representation is not a fundamental representation for $G = \mathrm{SU}(n)$ nor $\mathrm{Sp}(n)$:
The adjoint representation is given by the tensor product of $\textbf{n}$ and $\overline{\textbf{n}}$ representations for $\mathrm{SU}(n)$, and the degree two symmetric tensor product of the natural representation for $\mathrm{Sp}(n)$.
Therefore, the operator $\mathcal{O}_2$ is not generated by the fusion process if $k = 1$.
Hence, we should apply the operator $\mathcal{O}_1$ for the perturbative expansion to discuss the finite temperature behavior for $\mathrm{SU}(n)_1$ and $\mathrm{Sp}(n)_1$, while we should use the other operator $\mathcal{O}_2$ for the remaining cases.

For the complex fermion system, the total partition function in the presence of the perturbative term $\delta \mathsf{H}$ and the external $G$-field coupling is given by the path integral over the fermion field,
\begin{align}
    Z & = \exp \qty ( - \beta F(\lambda_\sigma,h) )
    \nonumber \\
    & =
    \int \mathcal{D} \psi \mathcal{D} \psi^\dag \,
    \exp \qty [ - \int_{-\beta/2}^{\beta/2} \dd{\tau} \int_{-L/2}^{L/2} \dd{x} 
    \qty( 
    \psi^\dag (\partial_\tau + \ii \partial_x) \psi - \lambda_\sigma \mathcal{O}_\sigma + \frac{h}{2 \pi} J^1(\tau, x)
    )]
\end{align}
where $L$ is the system size, and $\beta = T^{-1}$ is the inverse temperature.
Splitting the free energy into the bulk and the impurity contributions, $F = L f_\text{bulk} + f_\text{imp}$, we obtain
\begin{align}
    \ee^{-\beta f_\text{imp}}
    & = 
    \left<
    \exp \qty [ - \int_{-\beta/2}^{\beta/2} \dd{\tau} \int_{-L/2}^{L/2} \dd{x} 
    \qty( 
     - \lambda_\sigma \mathcal{O}_\sigma + \frac{h}{2 \pi} J^1(\tau, x)  
     )]
    \right>
    \, ,
\end{align}
where the average is taken with respect to the path integral with $(\lambda_\sigma, h) = (0,0)$.
We obtain the same correlator expression in the real fermion system.
We will perturbatively evaluate this path integral to compute the low temperature behavior of the thermodynamic quantities.

\subsubsection{Specific heat}

We consider the low temperature behavior of the specific heat.
From the operator $\mathcal{O}_1$, we obtain~\cite{Affleck:1990zd}
\begin{align}
    C_\text{imp} = \lambda_1 \frac{k \dim \mathfrak{g}}{3} \pi^2 T
    \, ,
    \label{eq:C_imp1}
\end{align}
which exhibits the linear dependence on the temperature, a.k.a., the FL property.
As discussed above, this result is applied only to the case with $k = 1$ for $G = \mathrm{SU}(n)$ and $\mathrm{Sp}(n)$:
\begin{align}
    C_\text{imp}
    =
    \begin{cases}
    \displaystyle
    \lambda_1 \frac{n^2 - 1}{3} \pi^2 T 
    & (\mathrm{SU}(n)_1) \\[.5em]
    \displaystyle
    \lambda_1 \frac{n(2n+1)}{3} \pi^2 T
    & (\mathrm{Sp}(n)_1) 
    \end{cases}
\end{align}

From the operator $\mathcal{O}_2$, we instead obtain the following expression:
\begin{align}
    C_\text{imp} = 
    \begin{cases}
    \displaystyle
    \lambda_2^2 \pi^{2 \Delta_\phi + 1} \Delta_\phi^2 
    \qty( 1 - 2 \Delta_\phi )
    \frac{\Gamma(1/2 - \Delta_\phi) \Gamma(1/2)}{\Gamma(1 - \Delta_\phi)} 
    \dim \mathfrak{g} \qty ( h^\vee + \frac{k}{2} ) 
    \, 
    T^{2 \Delta_\phi}
    & (\Delta_\phi < 1/2) \\[1em]
    \displaystyle
    \lambda_2^2 \pi^{2}     
    \dim \mathfrak{g} \qty ( h^\vee + \frac{k}{2} ) 
    \, T \log \qty( \frac{T_\text{K}}{T} )
    & (\Delta_\phi = 1/2) \\[1em]
    \displaystyle
    \frac{4 \lambda_2^2 \pi^2 \Delta_\phi \, T_\text{K}^{2 \Delta_\phi - 1}}{(2 \Delta_\phi + 1)(2 \Delta_\phi - 1)}
    \dim \mathfrak{g} \qty ( h^\vee + \frac{k}{2} ) 
    \, T
    & (\Delta_\phi > 1/2) 
    \end{cases}
    \label{eq:C_imp}
\end{align}
In the derivation of this expression, we use the formula,
\begin{align}
    \left< \,
    J_{-1}^a \phi^a(\tau_1,0) J_{-1}^b \phi^b(\tau_2,0)
    \, \right> 
    = 
    \frac{(h^\vee + k/2) \dim \mathfrak{g}}{\abs{[\tau_1 - \tau_2]}^{2(\Delta_\phi + 1)}}
    \qquad
    \text{with}
    \qquad
    [x] = \frac{\beta}{\pi} \sin \qty( \frac{\pi}{\beta} x )
    \, .
    \label{eq:[x]}
\end{align}
The remaining computation is parallel with the earlier analysis.
See, for example,~\cite{Affleck:1990iv,Kimura:2016zyv} for details.
Compared to the dimension of the adjoint primary operator~\eqref{eq:adj_dim}, we have $\Delta_\phi \lessgtr 1/2$ $\iff$ $h^\vee \lessgtr k$.
Then, focusing on the temperature dependence, it shows 
\begin{align}
    C_\text{imp} \propto
    \begin{cases}
    T^{2 \Delta_\phi} 
    & (\Delta_\phi < 1/2;\ h^\vee < k) \\
    T \log (T_\text{K} / T)
    & (\Delta_\phi = 1/2;\ h^\vee = k) \\
    T
    & (\Delta_\phi > 1/2;\ h^\vee > k) 
    \end{cases}
\end{align}
Therefore, for $h^\vee \le k$, we have the fractional power (logarithmic) dependence on the temperature, which is a specific feature of the NFL, while it shows the standard FL behavior for $h^\vee > k$, which is compatible with the contribution of the operator $\mathcal{O}_1$~\eqref{eq:C_imp1}.

\subsubsection{Susceptibility}

We can similarly compute the susceptibility.
From the operator $\mathcal{O}_1$, we obtain~\cite{Affleck:1990zd}
\begin{align}
    \chi_\text{imp} = \lambda_1 \frac{k(h^\vee + k)}{2}
    \, ,
\end{align}
which is indeed the FL behavior as expected.
As well as the specific heat, this result is applied only to the case with $k = 1$ for $G = \mathrm{SU}(n)$ and $\mathrm{Sp}(n)$:
\begin{align}
    \chi_\text{imp}
    =
    \begin{cases}
    \displaystyle
    \lambda_1 \frac{n + 1}{2}
    & (\mathrm{SU}(n)_1) \\[.5em]
    \displaystyle
    \lambda_1 \frac{n + 2}{2}
    & (\mathrm{Sp}(n)_1) 
    \end{cases}
\end{align}

From the operator $\mathcal{O}_2$, we obtain the following expression:
\begin{align}
    \chi_\text{imp} = 
    \begin{cases}
    \displaystyle
    \lambda_2^2 \pi^{2 \Delta_\phi - 1}
    \qty( \frac{1}{2} - \Delta_\phi )
    \frac{\Gamma(1/2 - \Delta_\phi) \Gamma(1/2)}{\Gamma(1 - \Delta_\phi)} 
    \qty ( h^\vee + \frac{k}{2} )^2
    \, 
    T^{2 \Delta_\phi - 1}
    & (\Delta_\phi < 1/2) \\[1em]
    \displaystyle
    2 \lambda_2^2 
    \qty ( h^\vee + \frac{k}{2} )^2 
    \, \log \qty( \frac{T_\text{K}}{T} )
    & (\Delta_\phi = 1/2) \\[1em]
    \displaystyle
    \frac{2 \lambda_2^2 \, T_\text{K}^{2 \Delta_\phi - 1}}{2 \Delta_\phi - 1}
    \qty ( h^\vee + \frac{k}{2} )^2 
    & (\Delta_\phi > 1/2) 
    \end{cases}
\end{align}
In this case, we use the formula
\begin{align}
    &
    \left< \,
    J^1(z_1) J^1(z_2) J_{-1}^a\phi^a(z_3) J_{-1}^{b} \phi^b(z_4)
    \, \right>_\text{connected}
    \nonumber \\ &
    =
    \qty(
    \frac{1}{[z_1 - z_4]^2 [z_2 - z_3]^2}
    + \frac{1}{[z_1 - z_3]^2 [z_2 - z_4]^2}
    )
    \frac{(h^\vee + k/2)^2}{[z_3 - z_4]^{2 \Delta_\phi}}
\end{align}
where the function $[x]$ is defined in \eqref{eq:[x]}.
See~\cite{Affleck:1990iv,Kimura:2016zyv} for details.
The temperature dependence is given by
\begin{align}
    \chi_\text{imp} \propto
    \begin{cases}
    T^{2 \Delta_\phi - 1} 
    & (\Delta_\phi < 1/2;\ h^\vee < k) \\
    \log (T_\text{K} / T)
    & (\Delta_\phi = 1/2;\ h^\vee = k) \\
    \text{const.}
    & (\Delta_\phi > 1/2;\ h^\vee > k) 
    \end{cases}
\end{align}
Then, for the NFL regime $h^\vee \le k$, it shows the fractional power (logarithmic) dependence as well as the specific heat, while it shows no temperature dependence for the FL regime $h^\vee > k$.

\subsubsection{Wilson ratio}

As seen in the analysis above, the thermodynamic quantities explicitly depend on the coupling constants $\lambda_{1,2}$, which are non-universal parameters of the model, although the temperature dependence seems universal.
In order to cancel such a coupling dependence, we consider the Wilson ration as a combination of the specific heat and the susceptibility:
\begin{align}
    R_\text{W} =
    \qty( \frac{\chi_\text{imp}}{C_\text{imp}} ) \Big/ \qty( \frac{\chi_\text{bulk}}{C_\text{bulk}} )
    \, .
\end{align}
From the perturbative analysis with the operator $\mathcal{O}_1$, we obtain the Wilson ratio as follows:
\begin{align}
    R_\text{W} =
    \frac{h^\vee + 1}{\dim \mathfrak{g}} \, c
    =
    \begin{cases}
    \displaystyle
    \frac{n}{n - 1} 
    & (\mathrm{SU}(n)_1)
    \\[.5em]
    \displaystyle
    \frac{2(n + 2)}{2n + 1}
    & (\mathrm{Sp}(n)_1)
    \end{cases}
\end{align}
where $c$ is the total central charge~\eqref{eq:c_tot}.
Together with the operator $\mathcal{O}_2$, we obtain the Wilson ratio for $h^\vee \le k$ as follows:
\begin{align}
    R_\text{W} =
    \frac{b n}{3}
    \frac{(h^\vee + k/2)(h^\vee + k)^2}{h^{\vee 2} \dim \mathfrak{g}} \, c
    =
    \begin{cases}
    \displaystyle
    \frac{(n + k/2)(n + k)^2}{3n(n^2 - 1)}
    & (\mathrm{SU}(n)_k) \\[1em]
    \displaystyle
    \frac{(n - 2 + k/2)(n - 2 + k)^2}{3 (n - 1) (n-2)^2 }
    & (\mathrm{SO}(n)_k) \\[1em]
    \displaystyle
    \frac{2}{3}
    \frac{(n + 2 + k/2)(n +1 + k)^2}{(n + 1)^2 (2n+1) }
    & (\mathrm{Sp}(n)_k) \\[1em]
    \end{cases}
\end{align}
where $b$ is the symmetry parameter defined in \eqref{eq:b_parameter}.
For $h^\vee > k$, on the other hand, the contributions from the operators $\mathcal{O}_1$ and $\mathcal{O}_2$ are compatible, so that we should take into account both of them.
This makes the Wilson ratio non-universal: 
It depends on the ratio of the coupling constants $\lambda_1$ and $\lambda_2$. 
This phenomenon, called the Fermi/non-Fermi mixing, occurs in particular in this parameter regime~\cite{Kimura:2016zyv}.

\section{EFG of Kondo effect}\label{sec:EFG}

We briefly mention the case with the exceptional group symmetry, $G = EFG$.
In this case, there is no systematic conformal embedding as in \eqref{eq:embedding1} and \eqref{eq:embedding_real}.
Hence, it may have a difficulty in the formulation of the Kondo problem in the sense discussed there.
Nevertheless, the analysis discussed in Sec.~\ref{sec:g_factor} and Sec.~\ref{eq:thermodynamics} is similarly applicable to the exceptional groups $G = EFG$.

One thing we should specifically clarify is the leading irrelevant operator discussed in Sec.~\ref{eq:therm_imp}.
In fact, since the adjoint representation is one of the fundamental representations for $G = EFG$, we should take the operator $\mathcal{O}_2$ as the leading irrelevant operator for $\forall k \ge 1$, which implies the NFL behavior for arbitrary $k \ge h^\vee$.%
\footnote{%
This argument is not applied to $E_8$ since the unique primary field of $\widehat{\mathfrak{e}}_{8,k}$ is the identity at level one $k = 1$.
}
In the case with $k < h^\vee$, on the other hand, it would be described as the FL.

\section{Discussion}

In this paper, we have formulated the Kondo problem with generic symmetry group $G = ABCD$, and studied its thermodynamic properties based on the BCFT analysis.
Although BCFT provides a powerful tool to study the non-perturbative aspects of the Kondo effect, it would be instructive to verify the results obtained in this paper with other methodologies, e.g., numerical renormalization group, Bethe ansatz analysis.
In particular, since the exact method based on the Bethe ansatz is formulated for the integrable system with generic $G$~\cite{Reshetikhin:1987bz}, the $G$-Kondo system could be similarly explored with the Bethe ansatz with inhomogeneity. 
See a review article \cite{Tsvelick:1983} for details.

Even though we have theoretically studied the Kondo system with generic symmetry group $G$, it would be important to examine how to realize such a generalized situation.
There are several realizations for the SU($n$) systems as mentioned earlier (Sec.~\ref{sec:intro}), where a similar analysis is also applicable~\cite{Kimura:2018vxj}.
For the SO($n$) Kondo system, the Majorana fermion realizes the quantum impurity in the spinor representation of SO($n$).
From this point of view, it would be also interesting to consider the impurity which transforms in other representations of SO$(n)$, e.g., the vector representation.
For the Sp($n$) theory, it might be possible to use a similar setup to the SO$(n)$ case:
Actually both cases are based on the conformal embedding of the real fermion as shown in Sec.~\ref{sec:real_embed}.
For example, one can use the isomorphism, $\mathrm{Sp}(2) = \mathrm{SO}(5)$.
In this case, the (four-dimensional) spinor representation of $\mathrm{SO}(5)$ is directly identified with the natural representation of Sp$(2)$ (See Sec.~\ref{sec:g_Sp}).
This suggests the realization of the Sp$(n)$ Kondo system with the Majorana impurities by imposing the symplectic Majorana condition.

\subsubsection*{Acknowledgments}

We would like to thank E. Kogan and A. Tsvelik for their kind correspondences.
We are also grateful to Sho Ozaki for the earlier collaboration, which leads to a lot of results presented in this paper.
In addition, we would appreciate useful comments by the anonymous referee.
This work has been supported in part by ``Investissements d'Avenir'' program, Project ISITE-BFC (No.~ANR-15-IDEX-0003), and EIPHI Graduate School (No.~ANR-17-EURE-0002).

\appendix

\section{Lie algebras}\label{sec:Lie_alg}


We summarize the properties of Lie algebras used in this paper.
See, for example, \cite{Yamatsu:2015npn} for detailed information about the Lie algebras.

In the CFT analysis, we often use the dual Coxeter number, which is given as the quadratic Casimir element of the adjoint representation,
\begin{align}
    f^{acd} f^{bcd} = h^\vee \, \delta^{ab}
    \, ,
    \label{eq:hvee}
\end{align}
where $f^{abc}$ is the structure constant of the Lie algebra $\mathfrak{g}$.
We remark that it agrees with the Coxeter number $h = h^\vee$ for $G = ADE$, which is related to the rank, and the dimension of $\mathfrak{g}$,
\begin{align}
    \dim \mathfrak{g} = n (h + 1) \stackrel{G = ADE}{=} n (h^\vee + 1)
    \, .
\end{align}
Therefore, the central charge~\eqref{eq:cc} of the level $k = 1$ affine Lie algebra for $\mathfrak{g} = \mathfrak{ade}(n)$ coincides with the rank,
\begin{align}
    c \qty[ \widehat{\mathfrak{ade}(n)}_1 ] = n
    \, ,
\end{align}
which implies the realization with $n$ complex fermions (Frenkel--Kac construction~\cite{Frenkel:1980rn}).

We summarize these properties in Tab.~\ref{tab:h_hvee}. 
We notice that the dual Coxeter number and the dimension of SO$(n)$ are in general given as
\begin{align}
    h^\vee = n - 2
    \, , \qquad
    \operatorname{dim} \mathrm{SO}(n) = \frac{1}{2}(n^2 - n)
    \, .
\end{align}

The Weyl vector is defined as
\begin{align}
    \rho = \frac{1}{2} \sum_{\alpha \in \Delta_+(\mathfrak{g})} \alpha
    \, ,
    \label{eq:Weyl_vec}
\end{align}
with the set of the positive roots of $\mathfrak{g}$ denoted by $\Delta_+(\mathfrak{g})$.
It is also possible to express the Weyl vector $\rho$ as the sum of the fundamental weights.
See~\cite{Fulton:2004} for details.
The Weyl vector is explicitly given as
\begin{subequations}
\begin{align}
    \mathrm{SU}(n) \ : \quad &
    \rho = \qty(
    \frac{n - 1}{2}, \frac{n - 3}{2},\ldots, - \frac{n - 3}{2}, - \frac{n - 1}{2}
    )
    \, , \\
    \mathrm{SO}(2n + 1) \ : \quad &    
    \rho = 
    \qty(
    n - \frac{1}{2}, n - \frac{3}{2}, \ldots, \frac{1}{2}
    )
    \, , \\
    \mathrm{Sp}(n) \ : \quad &
    \rho = \qty(
    n, n - 1, \ldots, 2, 1
    )
    \, , \\
    \mathrm{SO}(2n) \ : \quad &    
    \rho = 
    \qty(
    n - 1, n - 2, \ldots, 1, 0
    )
    \, .
\end{align}
\end{subequations}

\begin{table}[t]
    \centering
    \begin{tabular}{cc|ccc} \hline\hline
        Classification & Lie group $G$ & Coxeter $h$ & dual Coxeter $h^\vee$ & dimension \\\hline
        $A_n$ & SU($n+1$) & $n+1$ & $n+1$ & $n (n + 2)$ \\
        $B_n$ & SO($2n+1$) & $2n$ & $2n-1$ & $n(2n+1)$ \\
        $C_n$& Sp($n$)& $2n$ & $n+1$ & $n(2n+1)$ \\        
        $D_n$ & SO($2n$) & $2n-2$ & $2n-2$ & $n(2n-1)$ \\
        $E_{6,7,8}$ & -- & 12, 18, 30 & 12, 18, 30 & 78, 133, 248 \\
        $F_4$ & -- & 12 & 9 & 52 \\
        $G_2$ & -- & 6 & 4 & 14 \\\hline\hline
    \end{tabular}
    \caption{Properties of Lie algebras.}
    \label{tab:h_hvee}
\end{table}

\section{Character formula}\label{sec:character}

The character of the representation $R$ of the Lie algebra $\mathfrak{g}$ has several equivalent expressions.
We here present the determinantal formula for the character of the classical groups, $G = ABCD$, a.k.a., the Schur, symplectic Schur, and orthogonal Schur functions.

Let $X = (x_1,\ldots,x_n) \in (\mathbb{C}^\times)^{\operatorname{rk} \mathfrak{g}}$ be an element of the maximal Cartan torus of $G$. 
Then the character of the representation $R$ parametrized by a (half-)integer sequence $\lambda$ is given as follows:
\begin{align}
    \chi_{R(\lambda)} (X) = 
    \begin{cases}
    s_\lambda(X) & (\mathrm{SU}(n)) \\
    o^{(\text{o})}_\lambda(X) & (\mathrm{SO}(2n+1)) \\
    sp_\lambda(X) & (\mathrm{Sp}(n)) \\
    o^{(\text{e})}_\lambda(X) & (\mathrm{SO}(2n))     
    \end{cases}
\end{align}
where $s_\lambda(X)$, $o^{(\text{o/e})}_\lambda(X)$, and $sp_\lambda(X)$ are the Schur, orthogonal Schur, and symplectic Schur functions, defined as the ratio of the determinants~\cite{Fulton:2004}
(See also~\cite{Sinha:2000ap,Garcia-Garcia:2019uve}):
\begin{subequations}
\begin{align}
    s_\lambda(X) & = 
    \frac{\det_{1 \le i, j \le n} \qty( x_j^{\lambda_i + n - i} )}{\det_{1 \le i, j \le n} \qty( x_j^{n - i} )}
    \, , \\
    o^{(\text{o})}_\lambda(X) & =
    \frac{\det_{1 \le i, j \le n} \qty( x_j^{\lambda_i + n - i + \frac{1}{2}} - x_j^{-(\lambda_i + n - i + \frac{1}{2})} )}{\det_{1 \le i, j \le n} \qty( x_j^{n - i + \frac{1}{2}} - x_j^{-(n - i + \frac{1}{2})} )}
    \, , \\
    sp_\lambda(X) & =
    \frac{\det_{1 \le i, j \le n} \qty( x_j^{\lambda_i + n - i + 1} - x_j^{-(\lambda_i + n - i + 1)} )}{\det_{1 \le i, j \le n} \qty( x_j^{n - i + 1} - x_j^{-(n - i + 1)} )}
    \, , \\    
    o^{(\text{e})}_\lambda(X) & =
    \frac{\det_{1 \le i, j \le n} \qty( x_j^{\lambda_i + n - i} + x_j^{-(\lambda_i + n - i)} ) + \det_{1 \le i, j \le n} \qty( x_j^{\lambda_i + n - i} - x_j^{-(\lambda_i + n - i)} )}{\det_{1 \le i, j \le n} \qty( x_j^{n - i} + x_j^{-(n - i)} )}    
    \, .
\end{align}
\end{subequations}
The highest weight representation $R(\lambda)$ is parametrized by a sequence of (half-)integers, $\lambda = (\lambda_1,\ldots,\lambda_n)$:
\begin{subequations}
\begin{align}
    \mathrm{SU}(n) \ : \quad 
    & \lambda = (\lambda_1 \ge \lambda_2 \ge \cdots \ge \lambda_n \ge 0) \in \mathbb{Z}_{\ge 0}^n
    \, , \\
    \mathrm{SO}(2n+1) \ : \quad
    & \lambda = (\lambda_1 \ge \lambda_2 \ge \cdots \ge \lambda_n \ge 0) \in \mathbb{Z}_{\ge 0}^n \ \text{or} \ \qty(\mathbb{Z}_{\ge 0} + \frac{1}{2})^n
    \, , \\
    \mathrm{Sp}(n) \ : \quad 
    & \lambda = (\lambda_1 \ge \lambda_2 \ge \cdots \ge \lambda_n \ge 0) \in \mathbb{Z}_{\ge 0}^n
    \, , \\
    \mathrm{SO}(2n) \ : \quad
    & \lambda = (\lambda_1 \ge \lambda_2 \ge \cdots \ge  |\lambda_n| \ge 0) \in \mathbb{Z}_{\ge 0}^n \ \text{or} \ \qty(\mathbb{Z}_{\ge 0} + \frac{1}{2})^n
    \, .  
\end{align}
\end{subequations}
For $\mathrm{SU}(n)$ and $\mathrm{Sp}(n)$, $\lambda$ is a partition, a sequence of non-negative non-increasing integers.
The denominators of the Schur functions have the following expressions:
\begin{subequations}
\begin{align}
    \det_{1 \le i, j \le n} \qty( x_j^{n - i} )
    & = \prod_{i<j}^n (x_i - x_j)
    \, , \\
    \det_{1 \le i, j \le n} \qty( x_j^{n - i + \frac{1}{2}} - x_j^{-(n - i + \frac{1}{2})} )
    & = \prod_{i<j}^n
    ( x_i^{\frac{1}{2}} x_j^{\frac{1}{2}} - x_i^{-\frac{1}{2}} x_j^{-\frac{1}{2}} )
    ( x_i^{\frac{1}{2}} x_j^{-\frac{1}{2}} - x_i^{-\frac{1}{2}} x_j^{\frac{1}{2}} )
    \prod_{i=1}^n (x_i^{\frac{1}{2}} - x_i^{-\frac{1}{2}})
    \, , \\
    \det_{1 \le i, j \le n} \qty( x_j^{n - i + 1} - x_j^{-(n - i + 1)} )
    & = \prod_{i<j}^n
    ( x_i^{\frac{1}{2}} x_j^{\frac{1}{2}} - x_i^{-\frac{1}{2}} x_j^{-\frac{1}{2}} )
    ( x_i^{\frac{1}{2}} x_j^{-\frac{1}{2}} - x_i^{-\frac{1}{2}} x_j^{\frac{1}{2}} )
    \prod_{i=1}^n (x_i - x_i^{-1})   
    \, , \\
    \det_{1 \le i, j \le n} \qty( x_j^{n - i} + x_j^{-(n - i)} ) & = 2 \prod_{i<j}^n
    ( x_i^{\frac{1}{2}} x_j^{\frac{1}{2}} - x_i^{-\frac{1}{2}} x_j^{-\frac{1}{2}} )
    ( x_i^{\frac{1}{2}} x_j^{-\frac{1}{2}} - x_i^{-\frac{1}{2}} x_j^{\frac{1}{2}} )
    \, ,
\end{align}
\end{subequations}
which correspond to the Weyl denominator formula (also known as the generalized Vandermonde determinant).


\bibliographystyle{utphys}
\bibliography{conf}

\end{document}